\documentstyle[12pt]{article}
\newcommand{\beq}{\begin{equation}}
\newcommand{\eeq}{\end{equation}}
\newcommand{\bea}{\begin{eqnarray}}
\newcommand{\eea}{\end{eqnarray}}

\def\dalam{\vcenter {
    \hbox {\vrule height8pt width0.4pt depth0.0pt
	   \vrule height8pt width7.2pt depth-7.6pt
           \vrule height8pt width0.4pt depth0.0pt
           \kern-8pt
	   \vrule height0.4pt width8pt depth0.0pt
          }}}
\begin{document}
\setcounter{page}{0}
\topmargin 0pt
\oddsidemargin 5mm
\renewcommand{\thefootnote}{\fnsymbol{footnote}}
\newpage
\setcounter{page}{0}
\begin{titlepage}
\begin{flushright}
QMW 94-21\\
\end{flushright}
\begin{flushright}
hep-th/9407163
\end{flushright}
\vspace{0.5cm}
\begin{center}
{\large {\bf Towards $c=0$ flows}} \\
\vspace{1.8cm}
\vspace{0.5cm}
{\large W. A. Sabra$^1$, O. A. Soloviev$^2$ and S. Thomas$^2$}\\
\vspace{0.5cm}
{$^1$\em Department of Physics, Royal Holloway and Bedford New College,\\
University of London, Egham, Surrey, United Kingdom}\\
\vspace{0.5cm}
{$^2$\em Physics Department, Queen Mary and Westfield College, \\
Mile End Road, London E1 4NS, United Kingdom}\\
\vspace{0.5cm}
\renewcommand{\thefootnote}{\arabic{footnote}}
\setcounter{footnote}{0}
\begin{abstract}
{We discuss some implications of the gravitational dressing of the
renormalization
group for conformal field theories perturbed by relevant operators.
The renormalization group flows are
defined with respect to the dilatation
operator associated with the $J_0^{(0)}$ mode of the $SL(2,R)$ affine algebra.
We discuss the possibility of passing under the $c=25$ barrier along
renormalization group flows in some models.}
\end{abstract}
\vspace{0.5cm}
\centerline{July 1994}
 \end{center}
\end{titlepage}
\newpage
\section{Introduction}

Polyakov's fundamental string is presented as a certain two dimensional
conformal field theory (CFT) coupled to two dimensional quantum gravity
\cite{Polyakov-1}. One of the physical conditions that has to be implemented
in such theories is the vanishing of the total
Virasoro central charge of the system \cite{Polyakov-2}. However this condition
alone is not sufficient to pick out a single string model. In fact, there are a
large number of such non-critical
strings all of which obey the given physical criterion. It is widely
believed that string field
theory will provide us with new principals for selecting the true ground state
of string theory. Although this theory is still being developed we can
already make out some of its basic features.

One of the most important notions in string field theory ought perhaps to
be the independence of its action on any particular CFT describing a certain
string vacuum. At present, we know how the string field action can be
obtained in the vicinity of a given CFT \cite{Zwiebach}. It has been observed
that this action is in fact invariant under any truly marginal perturbation
of the given conformal background \cite{Sen-1},\cite{Sen-2}. Such perturbations
may amount to topology change in the corresponding target manifolds.
In particular, it has been shown now by performing
truly marginal deformations one can connect two string solutions corresponding
to two different Calabi-Yau manifolds \cite{Aspinwall}. This result elucidates
a
certain background independence of the string field action.

Of course in order to talk at all about the
vicinity of a CFT, one has to define the
meaning
of the distance in the space of models. In general, this is a difficult
question to answer.
However, there does exist such a
definition for marginal deformations \cite{Sen-1},
\cite{Sen-2}. One can think of the space of models as a space
parametrized by coupling constants, and
variations in the space of models amount to
variations of these coupling constants. It is well known that one can arrive at
a new conformal theory by doing relevant perturbations on a given one
\cite{Wilson}. Thus, it might be interesting to check the covariance of string
field theory under relevant perturbations. This requires the definition of a
metric and connection on the space of couplings (see e.g. \cite{Sonoda},
\cite{Dolan}).

By relevant perturbations we mean deformations of a
CFT by operators whose conformal dimensions are less than one. In string
field theory such a perturbed system is to be considered coupled to 2d gravity.
Because relevant perturbations break the conformal symmetry of the matter
system, the treatment of the perturbed model is different from the
case of
a CFT perturbed by (truly) marginal operators in the presence of 2d gravity.
This interaction between the quantum gravity and the nonconformal matter system
leads to an interacting theory which must be conformally invariant. There will
be
constraints that arise as a consequence of
the conformal symmetry in the interacting system.
However, in this conformal system there is no longer a sensible splitting into
conformal matter and $SL(2,R)$ invariant quantum gravity.
Therefore, the gravitational dressing becomes an even more nontrivial issue in
such theories.

The aim of the present paper is to investigate renormalization group flows in
CFT's perturbed by relevant operators in the presence of 2d gravity. We would
like to point out that the renormalization group in the presence of quantum
gravity has been extensively studied during the last few years
\cite{Sabra}-\cite{Tanii}. Our attention will focus on effects which occur
when the quantum gravity couples to the CFT perturbed by relevant quasimarginal
operators. We will show in the context of particular matter systems, that
flows under the $c=25$ barrier are possible. From
the point of view of the results obtained in \cite{Schmidhuber}, flows under
the $c=25$ barrier should result in changing minkowskian string solutions into
euclidean string solutions. We will argue that such flows do not contradict the
basic notions of two dimensional quantum field theory.

Instead of working in the conformal gauge, we shall use
the light cone gauge
in which there is a definition of the dilatation operator due to Polyakov et al
\cite{Polyakov-2},\cite{Knizhnik}. Scale transformations of the
renormalization group are
defined with respect to this operator within the perturbation expansion in the
coupling constant. From this, one can calculate the gravitational dressing
of the beta
function. Given this beta function we can obtain a generalization of the
Cardy-Ludwig formula for the difference between Virasoro central charges of
the matter at IR and UV fixed points of the
flow \cite{Cardy} in the presence of 2d gravity. Based on this formula one can
exhibit flows between $c>25$ and $c<25$ conformal models coupled to 2d gravity.

The paper is organized as follows. In section 2 relevant perturbations of CFT's
in the presence of 2d gravity are considered. In section 3 we explore the
effect
of the
dressing of the renormalization group due to 2d quantum gravity. Finally, in
the last
section we discuss the possibility of R.G. flows under the $c=25$ barrier. As
an example to illustrate this,
we investigate the nonunitary WZNW model on the group manifold $G$, with
$\dim G=25$.

\section{Relevant perturbations in the presence of quantum gravity}

Given a CFT one can decide whether or not there is an operator $O$ with
following properties: that its conformal dimension is less than one and that
the operator product expansion with itself obeys
\begin{equation}
O~O=[O]~+~[I]~+~...\end{equation}
where $I$ is the identity operator. The square brackets denote the
contributions
of the corresponding operators and their descendants, whereas dots stand for
operators with scaling dimensions greater than one. If such an operator exists
in the spectrum of the theory defined by an action
$S$, then one can consider another model with action
\begin{equation}
S(\epsilon)=S~-~\epsilon\int d^2x~O(x).\end{equation}
If the coupling $\epsilon$ is small, then $S(\epsilon)$ can be treated as a
perturbation of $S$. The two properties of the operator $O$ mentioned above,
guarantee the
renormalizability of the perturbed theory, (2.2).
When $\epsilon\ne0$, the theory
$S(\epsilon)$ is no longer conformal except when $O$ is a truly
marginal operator.

Now consider coupling $S(\epsilon)$ to 2d gravity. General coordinate
invariance allows us to set
two components of metric to fixed values, leaving a single dynamical component.
For our proposes it is convenient to chose the light cone gauge:
\begin{equation}
d^2s=dx^+dx^-~+~h_{++}(dx^+)^2.\end{equation}
Here $h_{++}$ is only quantum gravitational dynamical degree of freedom.

As usual, the interaction between the matter system $S(\epsilon)$ and the
gravity is through the former's stress tensor.
\begin{equation}
S(\epsilon;h)=S(\epsilon)~+~\int d^2x~h_{++}T_{--}.\end{equation}
When the operator $O$ has conformal dimension\footnote{This conformal dimension
is to be computed by taking into account gravitational dressing, i.e in
accordance with the KPZ formula \cite{Knizhnik}.}
$\lambda\ne1$, then the stress
tensor acquires a nonvanishing trace. To leading order in $\epsilon$ the trace
is given by the conservation equation
\begin{equation}
\nabla_+T_{--}~+~\partial_-\Theta=0,\end{equation}
where
\begin{equation}
\Theta=-2\epsilon~(1-\lambda)O.\end{equation}

There are two gauge constraints corresponding to the
two gauge conditions. Namely,
\begin{equation}
T_{++}={\delta\Gamma\over\delta h_{--}}=0,~~~~~~~h_{--}=0\end{equation}
and
\begin{equation}
{\delta\Gamma\over\delta h_{+-}}=
\Theta~-~\frac{c}{6}\partial^2_-h_{++}=0,~~~~~~~h_{+-}=1.\end{equation}
Here $\Gamma$ is the full effective action of 2d gravity \cite{Polyakov-3},
whereas $c$ is the Virasoro central charge of the CFT described by $S$. In the
light cone gauge, by
definition
\begin{equation}
e^{\frac{ic}{6}\Gamma}=\langle e^{i\int d^2x~h_{++}T_{--}}\rangle_{matt}.
\end{equation}

{}From eq. (2.9) one can obtain the equation of motion for $h_{++}$. To leading
order in $\epsilon$ one finds
\begin{equation}
\frac{c}{6}~\partial_-^3h_{++}~-~2\epsilon(1-\lambda)\partial_-O=0.
\end{equation}

The system of eqs. (2.7), (2.8) and (2.10) completely defines the quantum
theory (2.4) to leading order in the perturbation. In particular the
constraint given by eq. (2.8) leads to the conformal invariance of the theory
(2.4).

\section{Gravitational dressing of the renormalization group}

The renormalization group
beta function of the coupling $\epsilon$ to leading orders
in $\epsilon$ is given as follows (see, e.g. \cite{Klebanov})
\begin{equation}
\beta=(2-2\lambda)\epsilon~-~\pi C\epsilon^2~+~{\cal
O}(\epsilon^3),\end{equation}
where the constant $C$ is related to the coefficient of the three point
function
\begin{equation}
\langle\langle O(x_1)O(x_2)O(x_3)\rangle\rangle={C~||O||^2\over
x_{12}^{2\lambda}x_{13}^{2\lambda}x_{23}^{2\lambda}}\end{equation}
with $x_{12}=x_1-x_2$ etc. Here
\begin{equation}
||O||^2=\langle\langle O(1)O(0)\rangle\rangle,\end{equation}
where the double brackets are defined according to
\begin{equation}
\langle\langle\cdot\cdot\cdot\rangle\rangle=\int{\cal D}h_{++}
\langle\cdot\cdot\cdot ~e^{i\int d^2x~h_{++}T_{--}}\rangle_{matt}.
\end{equation}

The problem of gravitational dressing has been solved in \cite{Klebanov}, where
the Ward identities for dressed correlation functions have been derived. These
Ward identities allow one to compute Green functions of conformal operators in
the presence of 2d gravity. In the case under consideration, we need to compute
the two and three point functions of the perturbation $O$ in the presence of
the quantum gravity.

Following ref. \cite{Klebanov}, we present the perturbation operator $O$
as follows
\begin{equation}
O(x)={\sum_{AA'}}\lambda_{AA'}~J^A_-~J_+^{A'},\end{equation}
where the currents obey the following conservation law
\begin{equation}
({\cal K}+2)
\partial_+J_-^A=h_{++}\partial_-J_-^A~+~\lambda(\partial_-h_{++})J^A_-
\end{equation}
with a similar equation existing for $J_+^{A'}$.

The general covariance gives rise to the following Ward identity
\cite{Polyakov-3}\begin{equation}
\langle\langle h_{++}(y)O(x_1)\cdot\cdot\cdot O(x_n)\rangle\rangle =
{1\over{\cal K}+2}
{\sum_i}\left[{(y^--x^-_i)^2\over(y^++x^+_i)}{\partial\over\partial
x_i^-}~+~2\lambda {(y^--x^-_i)\over(y^+-x^+_i)}\right]\langle\langle
O(x_1)\cdot\cdot\cdot O(x_n)\rangle\rangle,\end{equation}
where the constant ${\cal K}$ is defined below. Eq. (3.16) along with eqs.
(3.15) and (3.17) give rise to the Klebanov-Kogan-Polyakov differential
equation \cite{Klebanov}.

At this point the concept of scaling dimension has to be carefully
defined. Since we treat the
theory $S(\epsilon)$ in perturbation theory, the scaling dimension $\lambda$ of
the perturbation $O$ coincides with its scaling dimension in the theory $S$
coupled to 2d gravity. The formula for anomalous dimensions is
\cite{Polyakov-2},\cite{Polyakov-3},\cite{Knizhnik}:
\begin{equation}
\lambda~-~\Delta_0 ={\lambda(\lambda -1)\over {\cal K}+2},
\end{equation}
where $\Delta_0$ is scaling dimension of $O$ without gravity, whereas the
constant ${\cal K}$ is given by the formula
\begin{equation}
{\cal K}~+~2=\frac{1}{12}~(c-13~+~\sqrt{(c-1)(c-25)}).\end{equation}

In the case of the quasimarginal operator $O$, the KPZ formula (3.18) gives
rise to the relation
\begin{equation}
2-2\lambda={{\cal K}+2\over{\cal K}+1}(2-2\Delta_0).\end{equation}

On the other hand, it has been shown that the three point function gets dressed
according to the formula \cite{Klebanov}
\begin{equation}
C={{\cal K}+2\over{\cal K}+1}~C_0,\end{equation}
where $C_0$ is the coefficient of the three point function before 2d gravity
was turned on.

Combining eqs. (3.20), (3.21) in formula (3.11), we arrive at the following
relation
\begin{equation}
\beta={{\cal K}+2\over{\cal K}+1}~\beta_0,\end{equation}
with $\beta_0$ being the renormalization group beta function computed without
gravity. This formula reveals dressing of the renormalization group in
the presence of gravity. So far, in the light cone gauge,
this effect has been discovered in
\cite{Klebanov} for marginal perturbation  (in the conformal gauge this effect
was discussed in \cite{Schmidhuber},\cite{Ambjorn}). The important point to be
made is that the coefficient in front of $\beta_0$ in eq. (3.22) stems from the
gravitational dressing of correlation functions computed at $\epsilon=0$.
Therefore, this coefficient does not depend on the perturbation parameter
$\epsilon$. Hence, it is not going to depend on the scale ($t$) at least to
the leading orders under consideration. Also it is
necessary to point out that eq. (3.22) is a first order equation with respect
to $t$-derivative. Correspondingly the dressed running coupling $\epsilon$ can
be derived by integrating eq. (3.22). It might be interesting to understand
how the same coupling can be obtained from the second order equation discussed
in \cite{Klebanov},\cite{Schmidhuber}.

It is clear from equation (3.22)
that all fixed points of the perturbed CFT remain
critical points in the presence of 2d gravity at the same values of the
coupling constant. However, the flow from the UV critical point to the IR
critical point undergoes some changes. In particular, the difference between
Virasoro central charges at IR and UV conformal points will be different from
the same quantity in the absence of gravity.

\section{c=25 barrier}

The effect of gravitational dressing of the renormalization beta function in
the
presence of 2d gravity gives rise to a new formula for the difference between
the Virasoro central charges at the IR and UV conformal points. This formula
can be thought of as a generalization of the Cardy-Ludwig formula
\cite{Cardy}. We find that in the presence of the gravity the difference is
given as follows
\begin{equation}
\Delta c=c_{IR}~-~c_{UV}=-\left({{\cal K}+2\over{\cal K}+1}\right){y_0^3\over
C_0^2}||O||^2= {({\cal K}+2)^2\over({\cal K}+1)({\cal K}+3)}\Delta c_0,
\end{equation}
where $y_0=2-2\Delta_0$, whereas $\Delta c_0=c_{IR0}-c_{UV0}$ is the difference
in the absence of gravity. While $c_{UV}=c_{UV0}$, $c_{IR}\ne c_{IR0}$. Hence,
at the IR conformal point the conformal matter must couple to the quantum
gravity essentially nonminimally. In other words, along the renormalization
group flow the character of interaction between conformal matter and 2d gravity
changes drastically. This may give some hints at understanding the issue of
strong
gravitational coupling.

As one can see there are two singularities emerging at ${\cal K}=-1$ and at
${\cal K}=-3$ in formula (4.23).
The first value of ${\cal K}$ corresponds to $c=25$; whereas ${\cal
K}=-3$ to $c=1$. It is a well known fact that Polyakov's quantization of 2d
gravity fails when $1<c<25$. However there are no physical obstruction for the
existence of the quantum gravity coupled to conformal matter with $c$ in this
interval. So, there should exist some way to overcoming this problem.
Renormalization group flows may shed some light on this. Indeed, according to
Zamolodchikov's c-theorem \cite{Zamolodchikov},
the Virasoro central charge may decrease along the
flow. In particular, the central charge which was slightly larger than 25 at
the UV conformal point may become less than 25 at the IR conformal point. Since
all critical points must remain fixed points in the presence of 2d gravity
there is at least a possibility that flows connecting them exist
even if the gravity is turned on.
Because the coefficient $({\cal K}+2)/({\cal K}+1)$ is very large when $c$ is
closed to 25, the difference $\Delta c$ becomes even larger in the presence of
gravity. Therefore, $c$ may become less than 25 along the flow. Of course, the
gravity gets changed itself along the flow, however within perturbation theory
these changes can be taken into account due to eqs. (2.7), (2.8), (2.10).

In order to justify this conjecture we need a CFT with Virasoro central charge
being slightly larger than 25. It turns out that the theory of such a kind is
the nonunitary WZNW model on the group manifold $G$. This model is
characterized by {\it negative} integer
level $k$. In the large $|k|$ limit, the Virasoro central charge of this theory
is given as follows
\begin{equation}
c=c_{UV}=\dim G~+~{c_V(G)\dim G\over |k|}~+~{\cal O}(1/k^2),\end{equation}
where $c_V(G)$ is the eigenvalue of the quadratic Casimir operator in the
adjoint representation of $G$.
Thus, when $\dim G=25$ we have a desired situation: $c$ is just slightly
greater than 25.

Certainly, the nonunitary WZNW model is a highly nontrivial theory. Because it
has states with negative norm, its proper definition is quite complicated.
To a certain extent
this theory is defined by its algebraic properties, i.e by its affine and
Virasoro symmetries. We
are not going to discuss in this paper all these problems.
What we can say, is that there definitely exist unitary representations of the
Virasoro symmetry in the nonunitary WZNW model \cite{Soloviev-1}. In
particular, there is an operator which satisfies the two properties discussed
at the beginning of section 2 and at the same time this operator provides a
unitary Virasoro representation with respect to the M\"ubious invariant vacuum.
This operator, in fact, is nothing but the kinetic term of the WZNW model.
Thus, one can define an appropriate perturbation on the nonunitary WZNW model
\cite{Soloviev-2}.

It has been established that the nonunitary WZNW model perturbed by its kinetic
term flows to the unitary WZNW model \cite{Soloviev-2}. The difference
$\Delta c_0$ is given by \cite{Soloviev-2}
\begin{equation}
\Delta c_0={2c_V(G)\dim G\over k}~+~{\cal O}(1/k^2).\end{equation}

One can estimate the coefficient in the expression for the dressed beta
function given by eq. (3.22). One finds
\begin{equation}
{{\cal K}+2\over{\cal K}+1}\approx \sqrt{|k|\over4c_V(G)}.\end{equation}
This gives rise to
\begin{equation}
\Delta c\approx-25\sqrt{c_V(G)\over |k|}.\end{equation}
Thus, when $|k|$ is very large we obtain
\begin{equation}
c_{IR}=c_{UV}~+~\Delta c\approx 25-25\sqrt{{c_V(G)\over |k|}}<25.\end{equation}
This amounts to a R.G. flow under the $c=25$ barrier, and so justifies our
claim that flows
under $c=25$ can exist. However, based on the above analysis we cannot as yet
reveal
the way the conformal matter couples to 2d quantum gravity under the barrier.
One possibility might be to compute correlation functions under the
barrier by using perturbation theory.

Although our conjecture that flows under the $c=25$ barrier has been justified
by consideration of a nonunitary WZNW model only,
certainly there may exist other CFT's
with appropriate properties near the barrier. Moreover, formula (4.23) holds
for any flow between fixed points which are close to each other. Therefore,
our conclusion about flows under the $c=25$ barrier is based only on
computations in 2d quantum field theory\footnote{In the string cosmology
approach to R.G. flows in 2d gravity described in \cite{Schmidhuber}, our
results would suggest that there should exist
cosmological
solutions admitting a change of space-time signature. However, at present
it is not known
whether such solutions actually exist.}.

As mentioned above, it might be that quantum gravity itself also can be
explored under the $c=25$ barrier by perturbation theory. Equations
(2.8) and (2.10) allow one to present $h_{++}$ as
\begin{equation}
h_{++}=J^+-2x^-J^0+(x^-)^2J^-~+{12\over
c}~\epsilon(1-\lambda){\partial_+^2\over\dalam^2}O.\end{equation}
Here $J^i$ are the $SL(2,R)$ currents of the quantum gravity when the
perturbation is switched off. One can use this representation of $h_{++}$ to
compute all
Ward identities to leading orders in the perturbation theory.
We hope to return to this in a future publication.

\par \noindent
{\em Acknowledgement}: We would like to thank I. Klebanov, N. E. Mavromatos,
C. Schmidhuber and
A. Tseytlin for useful
discussions.
O. A. would like to thank the PPARC and the European Commision Human
Capital and Mobility Programme for financial support and also the Theory
Division at CERN where part of this work was completed. The work of S. T. is
supported by the Royal Society. The work of W. S. is supported by
PPARC.

\end{document}